\documentclass[aps,pra,reprint,groupedaddress]{revtex4-1}% used
\usepackage{graphicx}% added
\usepackage{color}% added
\usepackage{bm}% added
\usepackage{comment}% added
\usepackage{amssymb, amsmath}% added
\begin{document}
\title{Simulated bifurcation assisted by thermal fluctuation}
\author{Taro Kanao}
\email[]{taro.kanao@toshiba.co.jp}
\author{Hayato Goto}
\affiliation{Frontier Research Laboratory, Corporate Research \& Development Center, Toshiba Corporation, 1, Komukai-Toshiba-cho, Saiwai-ku, Kawasaki 212-8582, Japan}
\date{\today}
\begin{abstract}
Various kinds of Ising machines based on unconventional computing have recently been developed for practically important combinatorial optimization. Among them, the machines implementing a heuristic algorithm called simulated bifurcation have achieved high performance, where Hamiltonian dynamics are simulated by massively parallel processing. To further improve the performance of simulated bifurcation, here we introduce thermal fluctuation to its dynamics relying on the Nos\'e-Hoover method, which has been used to simulate Hamiltonian dynamics at finite temperatures. We find that a heating process in the Nos\'e-Hoover method can assist simulated bifurcation to escape from local minima of the Ising problem, and hence lead to improved performance. We thus propose heated simulated bifurcation and demonstrate its performance improvement by numerically solving instances of the Ising problem with up to 2000 spin variables and all-to-all connectivity. Proposed heated simulated bifurcation is expected to be accelerated by parallel processing.
\end{abstract}
\pacs{}
\maketitle
\section*{Introduction}
Unconventional computing with special-purpose hardware devices for solving combinatorial optimization problems has attracted growing interest due to practical importance.
Many combinatorial optimization problems can be mapped onto finding ground states of Ising spin models~\cite{Barahona1982, Lucas2014}, which is referred to as the Ising problem.
Special-purpose hardware devices for the Ising problem are called Ising machines.
Ising machines utilizing natural phenomena have been developed, such as quantum annealers~\cite{Kadowaki1998, Farhi2000, Farhi2001, Das2008, Albash2018, Hauke2020} with superconducting circuits~\cite{Johnson2011}, coherent Ising machines with pulse lasers~\cite{Wang2013, Marandi2014, Inagaki2016, Yamamoto2017, Honjo2021}, oscillator-based Ising machines~\cite{Wang2017, Wang2019a, Chou2019, Mallick2020, Vaidya2022}, and other Ising machines with various systems, such as stochastic nanomagnets~\cite{Sutton2017}, gain-dissipative systems~\cite{Kalinin2018}, spatial light modulators~\cite{Pierangeli2019}, memristor Hopfield neural networks~\cite{Cai2020}, and spin torque nano-oscillators~\cite{Houshang2020, Albertsson2021}.

Ising machines have also been implemented with special-purpose digital processors~\cite{Yamaoka2016, Tsukamoto2017, Aramon2019, Okuyama2019, Yamamoto2021, Patel2020, Leleu2021a} using simulated annealing (SA)~\cite{Kirkpatrick1983} and other algorithms.
Among such algorithms, simulated bifurcation (SB) is a recently proposed heuristic algorithm~\cite{Goto2019}.
SB originates from numerical simulations of Hamiltonian dynamics with bifurcation that is a classical counterpart of quantum adiabatic bifurcation in nonlinear oscillators~\cite{Goto2016, Goto2019a}, which itself has been studied actively~\cite{Nigg2017, Puri2017, Zhao2018, Goto2018a, Kewming2020, Onodera2020, Goto2020b, Kanao2021, Goto2021b}.
Simulation-based approaches such as SB allow one to deal with dense spin-spin interactions with high precision, which might be challenging for physical implementations.
Also, the classical dynamics can be simulated efficiently, unlike quantum dynamics.
SB can be accelerated by parallel processing with, e.g., field-programmable gate arrays (FPGAs)~\cite{Goto2019, Tatsumura2019, Zou2020, Tatsumura2021}, because of its capability of simultaneous updating of variables.
Recently proposed variants of SB have achieved faster and more accurate optimization~\cite{Goto2021} than original SB.

To further improve the performance of SB, here we introduce thermal fluctuation to SB.
SA can yield high-accuracy solutions by modeling thermal fluctuation~\cite{Kirkpatrick1983}, while quantum annealing utilizes quantum fluctuation~\cite{Kadowaki1998}.
These fluctuations can assist escape from local minima of the Ising problem and lead to higher solution accuracy.

Our method is based on the Nos\'{e}-Hoover method~\cite{Nose1984, Hoover1985}, which enables simulations of Hamiltonian dynamics at finite temperatures~\cite{Leimkuhler2004}.
Unlike SA, the Nos\'{e}-Hoover method does not use random numbers, namely, is deterministic, and thus the simplicity of SB is preserved.
We find that a simplified dynamics with only a heating process can improve the performance of SB, where an ancillary dynamical variable in the Nos\'{e}-Hoover method is replaced by a constant.
We numerically demonstrate this improvement by solving instances of the Ising problem with up to 2000 spin variables and all-to-all connectivity, which corresponds to the Sherrington-Kirkpatrick (SK) model introduced in studies of spin glasses~\cite{Sherrington1975, Parisi1979, Parisi1993}.
The SK model has been widely used to measure the performance of Ising machines~\cite{Inagaki2016, Aramon2019, Honjo2021, Goto2019, Okuyama2019, Goto2021, Yamamoto2021, Leleu2021a, Oshiyama2022}.
Proposed heated SB is also suitable for massively parallel implementations with, e.g., FPGAs.

\section*{Results and discussion}
\noindent{\bf SB with thermal fluctuation.}
First, we briefly explain the Ising problem and SB.
The Ising problem is to find $N$ Ising spins ${s_i=\pm1}$ minimizing a dimensionless Ising energy,
\begin{eqnarray}
	E_{\rm Ising}&=&-\frac{1}{2}\sum^N_{i=1}\sum^N_{j=1}J_{ij}s_is_j,\label{eq_Ising}
\end{eqnarray}
where $J_{ij}$ represents the interactions between $s_i$ and $s_j$ (${J_{ij}=J_{ji}}$ and ${J_{ii}=0}$).
The SB has two latest variants, ballistic SB (bSB) and discrete SB (dSB)~\cite{Goto2021}.
Both bSB and dSB are based on the following Hamiltonian equations of motion,
\begin{eqnarray}
	\dot{x}_i&=&a_0y_i,\label{eq_pos}\\
	\dot{y}_i&=&-\left[a_0-a(t)\right]x_i+c_0f_i,\label{eq_mom}
\end{eqnarray}
where $x_i$ and $y_i$ are respectively the positions and momenta corresponding to $s_i$, the dots denote time derivatives, $a(t)$ is a control parameter, and $a_0$ and $c_0$ are constants.
The force due to the interactions, $f_i$, are given by
\begin{eqnarray}
	f_i&=&\sum_{j=1}^NJ_{ij}x_j,\hspace{3.5em}\text{for bSB,}\label{eq_fb}\\
	f_i&=&\sum_{j=1}^NJ_{ij}\mathrm{sgn}\left(x_j\right),\hspace{1em}\text{for dSB},\label{eq_fd}
\end{eqnarray}
where ${\mathrm{sgn}(x_j)}$ is the sign of $x_j$.
Time evolutions of $x_i$ are calculated by solving Eqs.~(\ref{eq_pos}) and (\ref{eq_mom}) with the symplectic Euler method~\cite{Leimkuhler2004}, where the positions $x_i$ are confined within ${\left|x_i\right|\leq1}$ by perfectly inelastic walls at ${x_i=\pm1}$, that is, if ${\left|x_i\right|>1}$ after each time step, $x_i$ and $y_i$ are set to ${x_i=\mathrm{sgn}\left(x_i\right)}$ and ${y_i=0}$.
With increasing $a(t)$ from zero to $a_0$, bifurcations to ${x_i=\pm1}$ occur, and the signs ${s_i=\mathrm{sgn}\left(x_i\right)}$ yield a solution to the Ising problem.
A solution at the final time is at least a local minimum of the Ising problem~\cite{Goto2021}.
Ballistic behavior in bSB leads to fast convergence to a local or approximate solution, while the discretized $f_i$ in dSB enable higher solution accuracy with a longer time.

Here we apply the Nos\'e-Hoover method~\cite{Nose1984, Hoover1985, Leimkuhler2004} with a finite temperature $T$ to Eqs.~(\ref{eq_pos}) and (\ref{eq_mom}), obtaining
\begin{eqnarray}
	\dot{x}_i&=&a_0y_i,\\
	\dot{y}_i&=&-\left[a_0-a(t)\right]x_i+c_0f_i-\xi y_i,\label{eq_NH1}\\
	\dot{\xi}&=&\frac{1}{M}\left(\sum_{i=1}^Ny_i^2-NT\right),\label{eq_NH}
\end{eqnarray}
where $\xi$ is an ancillary variable playing a role of thermal fluctuation, and $M$ is a parameter (mass).
The variable $\xi$ controls an instantaneous temperature defined by
\begin{eqnarray}
	T_{\rm inst}&=&\frac{1}{N}\sum_{i=1}^Ny_i^2,\label{eq_TInst}
\end{eqnarray}
to be a given $T$ as follows.
When $T_{\rm inst}$ is smaller than $T$, $\dot{\xi}$ is negative according to Eq.~(\ref{eq_NH}), which makes $\xi$ negative.
Then ${|y_i|}$ increase owing to the last term in Eq.~(\ref{eq_NH1}), and thus $T_{\rm inst}$ increases and approaches $T$, which can be regarded as heating.
To the contrary, when ${T_{\rm inst}>T}$, cooling occurs.

We found that SB gives better solutions when $\xi$ is kept negative by negative initial $\xi$ and large $M$ (leading to ${\dot{\xi}\simeq0}$).
This observation suggests that the heating can improve SB but the cooling is unnecessary.
This may be because increased ${\left|y_i\right|}$ by the heating can lead to escape from local minima of the Ising problem.
Furthermore, small ${|\dot{\xi}|}$ due to large $M$ above implies that constant $\xi$ can play a similar role, and then we found that $\xi$ replaced by a negative constant $-\gamma$ ${(\gamma>0)}$ can rather yield higher performance.
The constant $\gamma$ is regarded as a rate of the heating.

Thus, in this paper, we propose SB with a heating term, which we call heated bSB (HbSB) and dSB (HdSB), as follows,
\begin{eqnarray}
	\dot{x}_i&=&a_0y_i,\label{eq_PosHeat}\\
	\dot{y}_i&=&-\left[a_0-a(t)\right]x_i+c_0f_i+\gamma y_i.\label{eq_MomHeat}
\end{eqnarray}
We numerically solve Eqs.~(\ref{eq_PosHeat}) and (\ref{eq_MomHeat}) by discretizing the time by ${t_{k+1}=t_k+\Delta t}$ with a time interval $\Delta t$, and by calculating $x_i\left(t_{k+1}\right)$ and $y_i\left(t_{k+1}\right)$ from $x_i\left(t_k\right)$ and $y_i\left(t_k\right)$ in each time step.
Here note that the symplectic Euler method is not applicable for Eqs.~(\ref{eq_PosHeat}) and (\ref{eq_MomHeat}), because these equations are no longer Hamiltonian equations owing to the term $\gamma y_i$~\cite{Leimkuhler2004}.
We empirically found that solution accuracy can be improved by the same update as previous bSB and dSB~\cite{Goto2021} followed by an update corresponding to the term $\gamma y_i$.
This ordering results in nonzero momenta by the heating and can prevent from getting stuck at the walls.
See Methods for a detailed algorithm.

In the following, we compare heated SB with previous SB by solving instances of the Ising problem with all-to-all connectivity, where $J_{ij}$ are randomly chosen from $\pm1$ with equal probabilities (corresponding to the SK model).
The control parameter $a(t)$ is linearly increased from 0 to $a_0$.
The constant parameters are set as ${a_0=1}$ and
\begin{eqnarray}
	c_0&=&\frac{c_1}{\sqrt{N}},
\end{eqnarray}
where $c_1$ is a parameter tuned around 0.5, which is based on random matrix theory~\cite{Goto2019, Goto2021}.
$x_i$ and $y_i$ are initialized by uniform random numbers in the interval ${(-1, 1)}$.
\\

\noindent{\bf Performance for a 2000-spin Ising problem.}
We first solve a benchmark instance called K$_{2000}$, which is a 2000-spin instance of the Ising problem with all-to-all connectivity~\cite{Goto2019, Goto2021, Inagaki2016, Yamamoto2021}.
K$_{2000}$ is often expressed as a MAX-CUT problem.
The MAX-CUT problem is given by weights $w_{ij}$ with ${w_{ij}=w_{ji}}$, and the following cut value $C$ is maximized,
\begin{eqnarray}
	C&=&\frac{1}{2}\sum_{i=1}^N\sum_{j=1}^Nw_{ij}\frac{1-s_is_j}{2}\\
	&=&-\frac{1}{2}E_{\rm Ising}-\frac{1}{4}\sum_{i=1}^N\sum_{j=1}^NJ_{ij},\label{eq_Cut}
\end{eqnarray}
where in Eq.~(\ref{eq_Cut}), $C$ has been related to $E_{\rm Ising}$ [Eq.~(\ref{eq_Ising})] by ${w_{ij}=-J_{ij}}$.
Thus the MAX-CUT problem can be reduced to the Ising problem.

\begin{figure}
	\includegraphics[width=8.5cm]{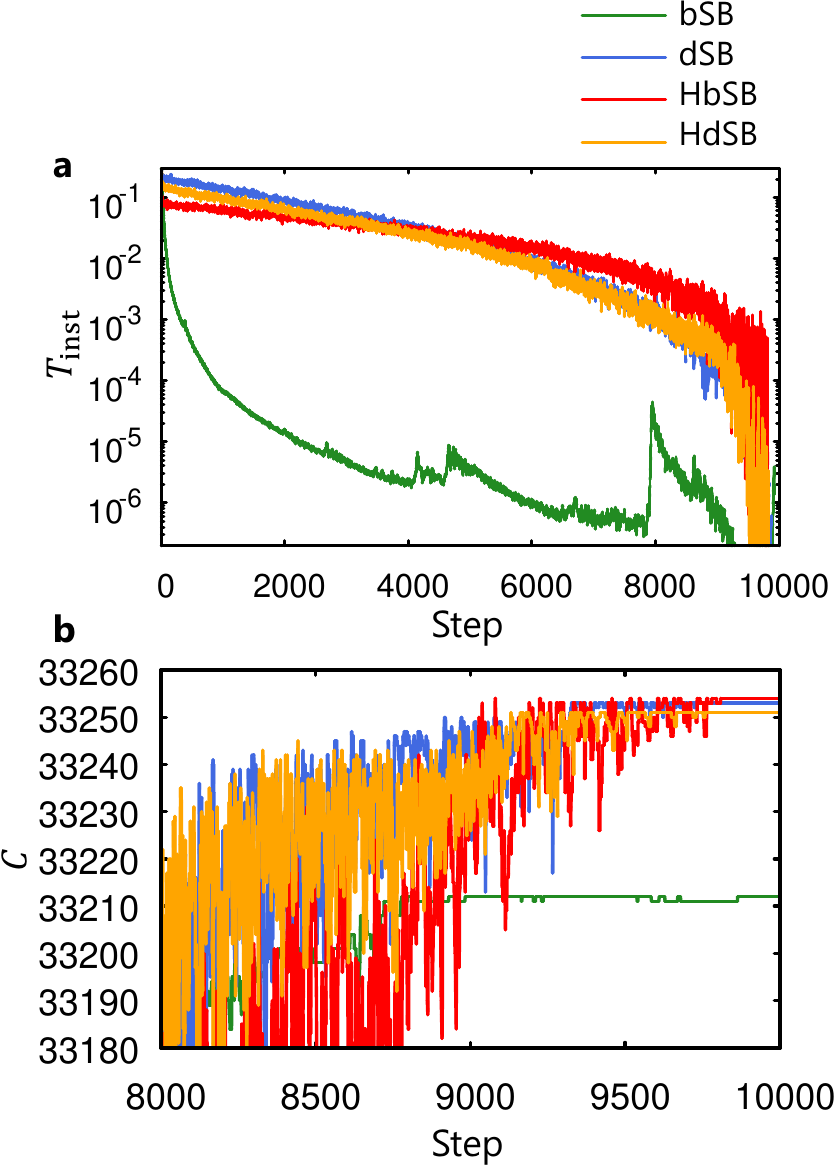}%
	\caption{{\bf Time evolutions in simulated bifurcation (SB) for a 2000-spin Ising problem (K$_{2000}$).}
			{\bf a} Instantaneous temperatures.
			{\bf b} Instantaneous cut values.
			Here, bSB, dSB, HbSB, and HdSB denote ballistic SB, discrete SB, heated bSB, and heated dSB, respectively.
			Time is represented by time steps.
			Parameters are ${\Delta t=0.7}$ and ${c_1=0.6}$ for bSB, ${\Delta t=1.1}$ and ${c_1=0.6}$ for dSB, ${\Delta t=1.1}$, ${c_1=0.9}$ and ${\gamma=0.5}$ for HbSB, and ${\Delta t=1.1}$, ${c_1=0.7}$ and ${\gamma=0.06}$ for HdSB.
		\label{fig_TInstC}}
\end{figure}
To confirm the effect of the heating, we calculate the instantaneous temperature $T_{\rm inst}$ in Eq.~(\ref{eq_TInst}) and the cut value $C$ in Eq.~(\ref{eq_Cut}) at every time step.
Figure~\ref{fig_TInstC}a shows typical examples of time evolutions of $T_{\rm inst}$.
Here parameters are the ones optimized in advance, which are explained later.
For bSB, $T_{\rm inst}$ rapidly decreases owing to collisions with the perfectly inelastic walls, while $T_{\rm inst}$ for HbSB is kept much higher owing to the heating, as expected.
For dSB, $T_{\rm inst}$ is also higher than bSB, because the discretized forces $f_i$ in Eq.~(\ref{eq_fd}) increase energy, violating conservation of energy~\cite{Goto2021}.
In comparison between HbSB and dSB, $T_{\rm inst}$ for HbSB is higher than dSB around the end of the evolution.
HdSB shows similar $T_{\rm inst}$ to dSB, because the optimal rate of the heating $\gamma$ for HdSB is small.

Figure~\ref{fig_TInstC}b shows last parts of typical time evolutions of $C$.
For bSB, $C$ is almost constant for the last 1000 time steps, while $C$ for HbSB continues to fluctuate until nearly the end of the evolution owing to the heating.
The fluctuation for HbSB looks similar to the fluctuations for dSB and HdSB.

\begin{figure}
	\includegraphics[width=8.5cm]{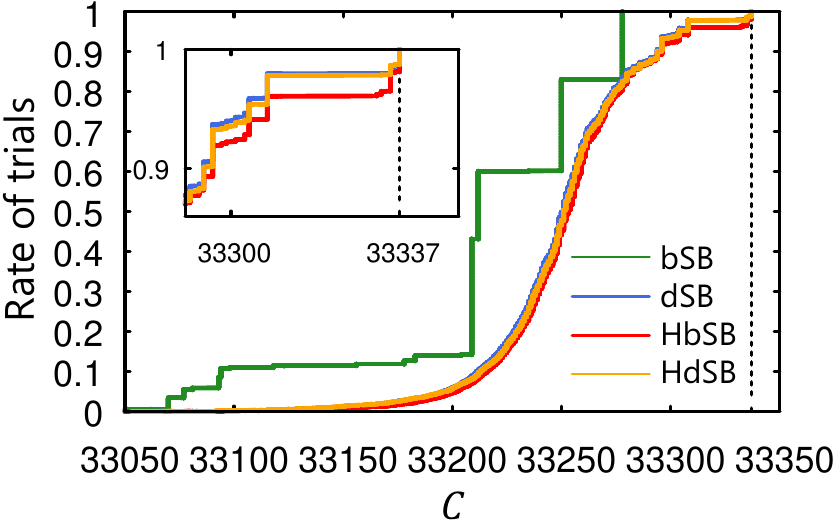}%
	\caption{{\bf Cumulative distributions of cut values for K$_{2000}$.}
			The dashed vertical line indicates the best known cut value 33337.
			Inset shows a magnification around the best known cut value.
			Here the number of time steps is ${N_{\rm s}=10000}$, and parameters $\Delta t$, $c_1$, and $\gamma$ are the same as in Fig.~\ref{fig_TInstC}.
			SB means simulated bifurcation, and b, d, and H denote ballistic, discrete, and heated, respectively.
		\label{fig_dist}}
\end{figure}
Next, we solve K$_{2000}$ in 10$^4$ trials with random initial $x_i$ and $y_i$.
In one trial, $C$ is evaluated every 100 time steps, and the best value is output.
Figure~\ref{fig_dist} shows examples of cumulative distributions of $C$, where the number of trials giving cut values lower than $C$ are normalized by the total number of trials, 10$^4$.
For bSB, the majority of trials results in one of a few values of $C$ between 33200 and 33300.
On the other hand, dSB, HbSB, and HdSB yield broader distributions, owing to fluctuations (higher instantaneous temperatures) in their dynamics.
It is notable that the heating makes the distribution of bSB similar to dSB (and HdSB).
Inset in Fig.~\ref{fig_dist} shows a magnification around the best known cut value, ${C_{\rm best}=33337}$~\cite{Goto2021}.
The distributions for HbSB and HdSB are shifted to larger $C$ than dSB, indicating improvement by the heating.

We then evaluate average and maximum $C$ in the $10^4$ trials and probability $P$ for obtaining $C_{\rm best}$.
Here, $P$ is estimated by dividing the number of trials obtaining $C_{\rm best}$ by the total number of trials.
Also, using $P$ and the number of time steps, $N_{\rm s}$, we calculate the number of time steps required to find $C_{\rm best}$ with a probability of 99\%, which we call step-to-solution $S$, given by
\begin{eqnarray}
	S&=&N_{\rm s}\frac{\log0.01}{\log(1-P)}.
\end{eqnarray}
Step-to-solution is a useful measure of performance of an algorithm~\cite{Leleu2021a}.
(Time-to-solution is often used to measure performance of Ising machines~\cite{Aramon2019, Goto2021, Leleu2021a, Patel2020}, but it depends on not only algorithms but also implementations to hardware devices.
Time-to-solution equals step-to-solution multiplied by a computation time for one time step.)
Here the parameters $\Delta t$, $c_1$, and $\gamma$ are set such that $S$ are minimized.
For bSB, instead, average $C$ is maximized, because we found $P=0$ for bSB and could not estimate $S$.

\begin{figure}
	\includegraphics[width=8.5cm]{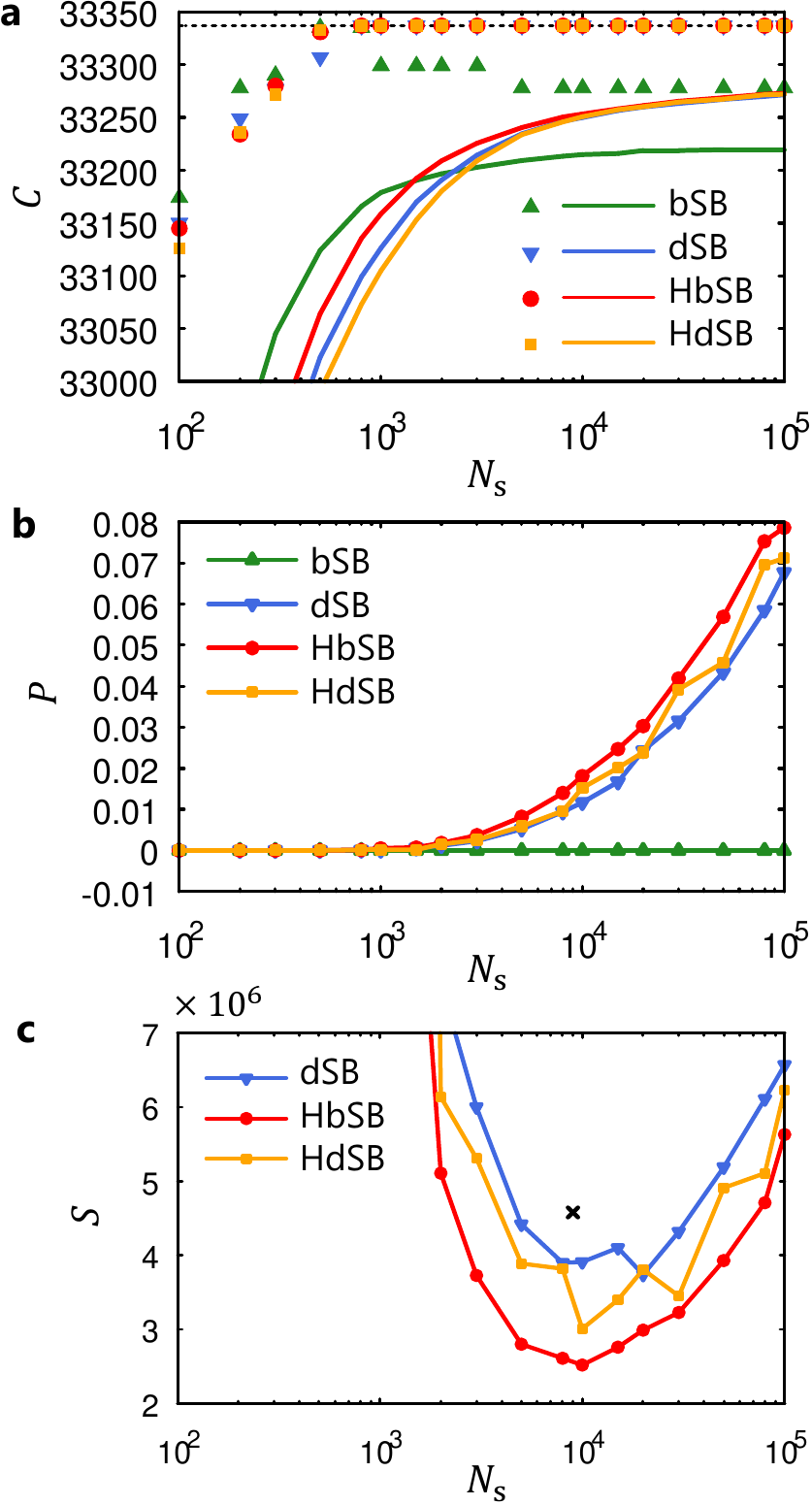}%
	\caption{{\bf Comparison of performance for K$_{2000}$.}
			{\bf a} Average (lines) and maximum (symbols) cut values $C$ as functions of the number of time steps, $N_{\rm s}$.
			The horizontal dashed line indicates the best known cut value ${C_{\rm best}=33337}$.
			The maximum $C$ for dSB, HbSB, and HdSB coincide with $C_{\rm best}$ for ${N_{\rm s}\geq800}$.
			{\bf b} Probabilities for obtaining $C_{\rm best}$.
			{\bf c} Step-to-solution.
			The black cross is step-to-solution obtained from previously reported data for dSB~\cite{Goto2021}.
			Parameters $\Delta t$, $c_1$, and $\gamma$ here are the same as in Fig.~\ref{fig_TInstC}.
			SB means simulated bifurcation, and b, d, and H denote ballistic, discrete, and heated, respectively.
		\label{fig_CPS}}
\end{figure}
Figure~\ref{fig_CPS}a shows average and maximum $C$ as functions of $N_{\rm s}$.
For large $N_{\rm s}$, average $C$ for dSB, HbSB, and HdSB are larger than that for bSB.
$C_{\rm best}$ is reached by three SBs other than bSB.
Figure~\ref{fig_CPS}b shows that, for large $N_{\rm s}$, $P$ are the highest for HbSB, followed by HdSB, dSB, and bSB in this ordering.

Figure~\ref{fig_CPS}c shows $S$ as functions of $N_{\rm s}$.
Each SB has a minimum of $S$ at certain $N_{\rm s}$, and in the following we compare $S$ minimized with respect to $N_{\rm s}$.
The black cross represents a value obtained from previously reported data for dSB~\cite{Goto2021}.
The present value of $S$ for dSB is smaller than the previous value, because in this study the parameters $\Delta t$ and $c_1$ are optimized for K$_{2000}$ while not in the previous study.
In Fig.~\ref{fig_CPS}c, at optimal $N_{\rm s}$, $S$ for HbSB is the smallest.
Compared with dSB, $S$ are reduced by 32.7\% for HbSB and 19.7\% for HdSB.
These results demonstrate that the heating improves the performance.
Although dSB performs better than bSB, bSB is much more improved by the heating than dSB, and resulting HbSB shows higher performance than HdSB.
\\

\noindent{\bf Performance for 100 instances of a 700-spin Ising problem.}
Finally we examine the performance for instances other than K$_{2000}$ by solving 100 instances of the Ising problem with 700 spin variables and all-to-all connectivity~\cite{Goto2021, Leleu2021a}.
For these instances, reference solutions for estimating step-to-solution were obtained by SA with sufficiently long annealing times and many iterations, which are expected to be close to optimal solutions~\cite{Goto2021}.
Here each instance is solved in $10^4$ trials, and $C$ (or $E_{\rm Ising}$) is evaluated at the last time step.
We set the parameters to the values optimized in K$_{2000}$.

\begin{figure}[b]
	\includegraphics[width=8.5cm]{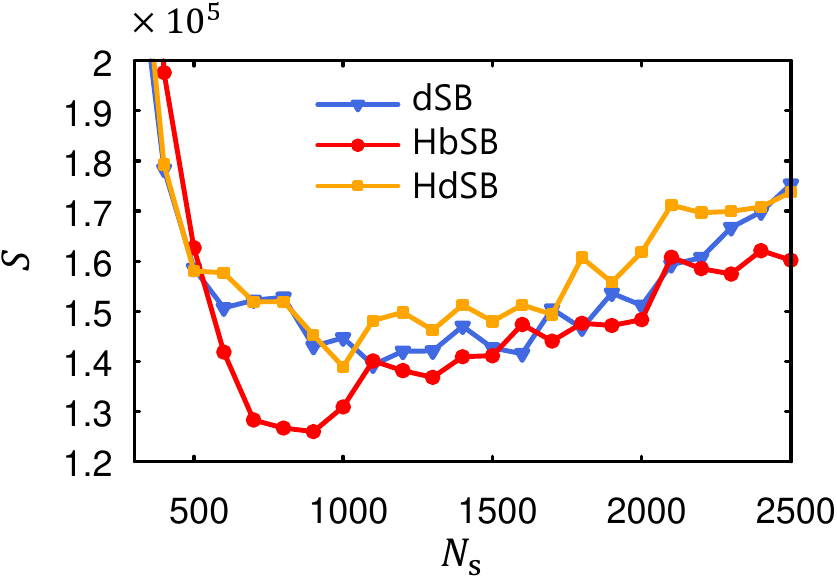}%
	\caption{{\bf Comparison of performance for 100 instances of a 700-spin Ising problem.}
			The medians of step-to-solution for the 100 instances are shown as functions of the number of time steps, $N_{\rm s}$.
			Parameters $\Delta t$, $c_1$, and $\gamma$ are the same as in Fig.~\ref{fig_TInstC}.
			SB means simulated bifurcation, and b, d, and H denote ballistic, discrete, and heated, respectively.
		\label{fig_MedSTS}}
\end{figure}
Figure~\ref{fig_MedSTS} shows the medians of $S$ for the 100 instances~\cite{Goto2021, Leleu2021a} as functions of $N_{\rm s}$.
HbSB results in the smallest $S$ among four SBs ($S$ by bSB is much larger than those by the others).
In comparison with dSB, HbSB reduces $S$ by 9.55\%.
This result demonstrate that HbSB can improve the performance for not only K$_{2000}$ but also the other instances.

\section*{Conclusion}
We have demonstrated that SB can be improved by introducing fluctuation with a heating term, which has been obtained by replacing an ancillary dynamical variable in the Nos\'e-Hoover method by a constant rate of heating.
This heating can be effective for both bSB and dSB, but the improvements are much larger for bSB.
We have solved all-to-all connected 2000-spin and 700-spin instances of the Ising problem (the SK model) and have found that HbSB gives better step-to-solution than bSB, dSB, and HdSB.
Since proposed heated SB shares the simple dynamics with previous SB, we expect that heated SB will be accelerated by massively parallel processing implemented by, e.g., FPGAs.
This study also implies that further improvements of SB will be possible by simple physics-inspired modifications like the heating term introduced here.

\section*{METHODS}
\noindent{\bf Heated simulated bifurcation.}
First, the symplectic Euler method~\cite{Leimkuhler2004} is formally applied to the terms other than $\gamma y_i$ in Eqs.~(\ref{eq_PosHeat}) and (\ref{eq_MomHeat}),
\begin{eqnarray}
	\tilde{y}_i&=&y_i\!\left(t_k\right)\!+\!\left\{\!-\!\left[a_0\!-\!a\!\left(t_k\right)\right]\!x_i\!\left(t_k\right)\!+\!c_0f_i\right\}\!\Delta t,\label{eq_SyEuY}\\
	\tilde{x}_i&=&x_i\!\left(t_k\right)\!+\!a_0\tilde{y}_i\Delta t,\label{eq_SyEuX}
\end{eqnarray}
where $f_i$ are calculated from ${x_i\!\left(t_k\right)}$ with Eqs.~(\ref{eq_fb}) and (\ref{eq_fd}), and the variables with the tildes denote temporary variables used within a time step.
Then, the perfectly inelastic walls work as
\begin{eqnarray}
	x_i\!\left(t_{k+1}\right)&=&g\!\left(\tilde{x}_i\right),\label{eq_WallX}\\
	\tilde{\tilde{y}}_i&=&h\!\left(\tilde{x}_i, \tilde{y}_i\right),\label{eq_WallY}
\end{eqnarray}
where the functions $g(x)$ and $h(x, y)$ are given by
\begin{eqnarray}
	g(x)&=&\left\{\begin{array}{cc}
		x,&|x|\leq1,\\
		1,&x>1,\\
		-1,&x<-1,
	\end{array}\right.\\
	h(x, y)&=&\left\{\begin{array}{cc}
		y,&|x|\leq1,\\
		0,&|x|>1.
	\end{array}\right.
\end{eqnarray}
Finally, we include the heating term, referring to the usual Euler method, as
\begin{eqnarray}
	y_i\!\left(t_{k+1}\right)&=&\tilde{\tilde{y}}_i\!+\!\gamma y_i\!\left(t_k\right)\!\Delta t.\label{eq_EuHeat}
\end{eqnarray}
Equations~(\ref{eq_SyEuY})--(\ref{eq_WallY}) and (\ref{eq_EuHeat}) are numerically solved, where the variables are represented as single precision floating-point real numbers.

\begin{acknowledgements}
We thank K. Tatsumura, R. Hidaka, Y. Hamakawa, M. Yamasaki, and Y. Sakai for valuable discussion.
\end{acknowledgements}

\end{document}